# GERBERTO E LA GEOGRAFIA TOLEMAICA

COSTANTINO SIGISMONDI

Università di Roma "La Sapienza" e ICRA
International Center for Relativistic Astrophysics
e-mail: sigismondi@icra.it

**Abstract:** The *horologia* are ancient ephemerides, tables of durations of the days and nights during all months of the year. The algorithms used to compute the *horologia* are here presented and the results are compared with the tables computed by Gerbert in the letter to Adam. The Ptolemaic geography is the framework in which Gerbert made his calculations.

## 1. Gerberto Scienziato Papa

Gerbert d'Aurillac è stato l'uomo più colto del suo tempo. Visse nel X secolo, francese, fu docente di retorica e delle scienze del quadrivio alla scuola cattedrale di Reims. Introdusse in Europa l'astrolabio e le cifre arabe escluso lo zero dalla Catalogna, dove aveva studiato. Segretario dell'Arcivescovo Adalberone e del re Ugo Capeto, Abate a Bobbio, in contatto con i sapienti ed i regnanti di tutta Europa, fu maestro di retorica e musica di Ottone III e da lui chiamato alla sede vescovile di Ravenna e poi a Roma nel 999. Fu Papa con il nome di Silvestro II fino al 12 maggio 1003. Sotto il suo pontificato la Chiesa giunse ufficialmente in Ungheria e Polonia, spinta dal suo ideale di costruire l'Europa su basi cristiane. La sua autorità morale e culturale ha influenzato l'establishment del quadrivio (Aritmetica, Musica, Geometria e Astronomia) nel curriculum studiorum delle prime Università. Quest'anno ricorre il millenario dalla sua morte, ed è stato organizzata una giornata di studi per ricordarlo.[1] In questo articolo si sperimentano con l'aiuto della trigonometria e del computer le difficoltà tecniche implicate dai problemi di *Geografia Astronomica* trattati da Gerberto con il suo allievo Adamo.

## 2. Gerberto e la Geografia

Di Gerberto ci sono pervenuti alcuni testi, la cui caratteristica è la stringatezza. Essi riguardano l'Astrolabio, le proporzioni delle Canne d'Organo, la retorica, la redazione degli atti del Concilio di Saint-Basle ed un vasto edd eclettico epistolario. Si tratta di un caso unico nel suo genere, e proprio grazie a Gerberto abbiamo un punto di vista autorevole della vita politica e culturale del X secolo, altrimenti troppo avvolta nel mistero e nel pregiudizio.
Esaminiamo qui la lettera spedita al fratello (in Cristo) Adamo, la numero 161 nella cronologia della studiosa americana Harriet Pratt Lattin.[2] La lettera è datata 10 marzo 989, un mese e mezzo dopo la morte dell'Arcivescovo Adalberone, di cui Gerberto diventerà successore di lì a qualche anno. Nella lettera precedente, del 7 marzo a Remi, monaco di Trier, ricorda questo evento e lo redarguisce per la sua insistenza di chiedergli la realizzazione di una sfera in tempi così complicati.
In altre lettere Gerberto parla di Globi e Sfere Celesti, di cui era un esperto costruttore e che poi usava come strumento didattico. Inoltre sappiamo che conosceva i problemi di astronomia computazionale (come il sorgere e tramontare degli astri) visto che sapeva usare l'astrolabio.

---

[1] Gerberto Scienziato e Papa, Pontificia Università Lateranense, 12 maggio 2003 http://141.108.24.98/eclisse/gerbert
[2] Harriet PRATT LATTIN, *The Letters of Gerbert with his papal privileges as Sylvester II*. Translated with an Introduction by Harriet PRATT LATTIN, Columbia University Press, New York, 1961, p. 189-191.

Ad Adamo, suo allievo a Reims, invia due *horologia*, ovvero tavole sulla lunghezza del giorno nei vari mesi dell'anno, tarate su due latitudini differenti: quella dell'Ellesponto e quella per la quale il giorno più lungo dura 18 ore (circa la latitudine di Stoccolma).
In questo studio vediamo come calcolare queste tavole, usando la trigonometria, evidenziando l'originalità del contributo di Gerberto rispetto ai testi di riferimento in uso all'epoca.

### 3. Globi e durata massima del dì

Illuminando un globo in modo che il polo Nord risulti in luce, e l'asse Nord Sud sia inclinata di ε=23°27' rispetto al piano del terminatore, si riproduce la situazione il giorno del Solstizio d'Estate. A causa dell'inclinazione ε sull'eclittica, il polo nord è tutto illuminato così come altre latitudini polari fino a λ=90°- ε, che è il circolo polare artico. Lì il giorno più lungo dura tutte e 24 le ore.

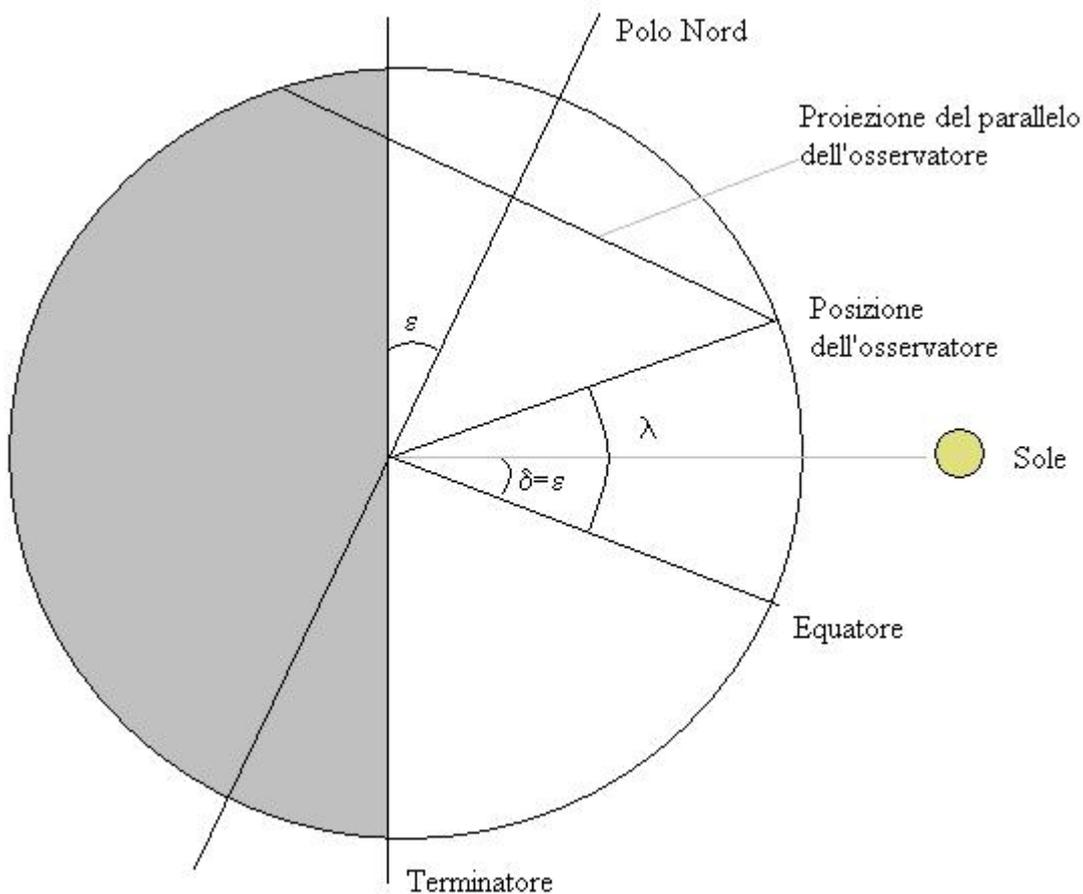

FIGURA 1

**Geometria per il calcolo della durata del dì più lungo.**

Proiezione sul piano laterale tangente al terminatore della sfera terrestre.
Il Sole è al solstizio estivo per l'emisfero Nord (tropico del Cancro), per cui la sua declinazione δ=ε

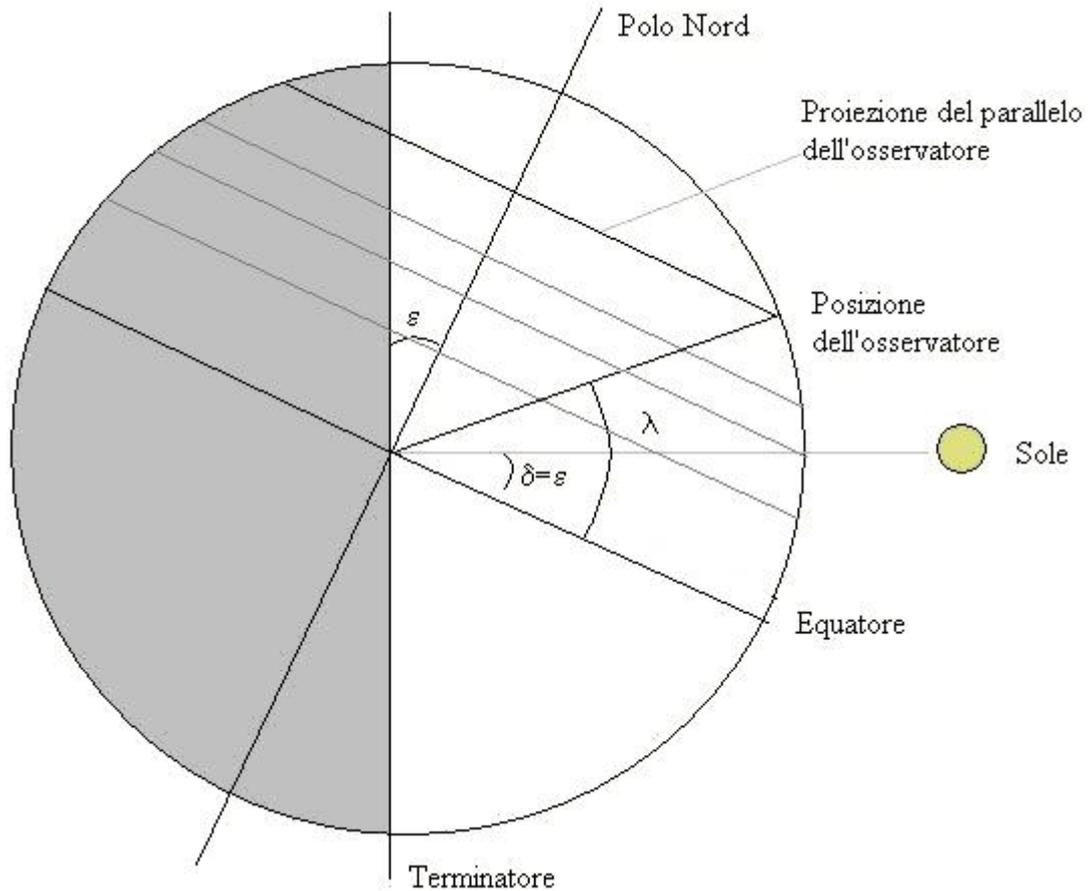

.

FIGURA 2

**Durata del dì a diverse latitudini.**

Sempre in proiezione sul piano tangente al terminatore vediamo i cerchi di eguale latitudine. Per conoscere la durata massima del dì per le latitudini comprese tra 90° -ε e l'equatore basta calcolare la frazione di cerchi di latitudine illuminata, e riportarla a 24 ore.
Dalla figura è immediatamente evidente che all'equatore tale frazione è pari esattamente alla metà. Tra l'equatore ed il circolo polare abbiamo tutti i casi intermedi tra 12 e 24 ore di massima lunghezza del dì.

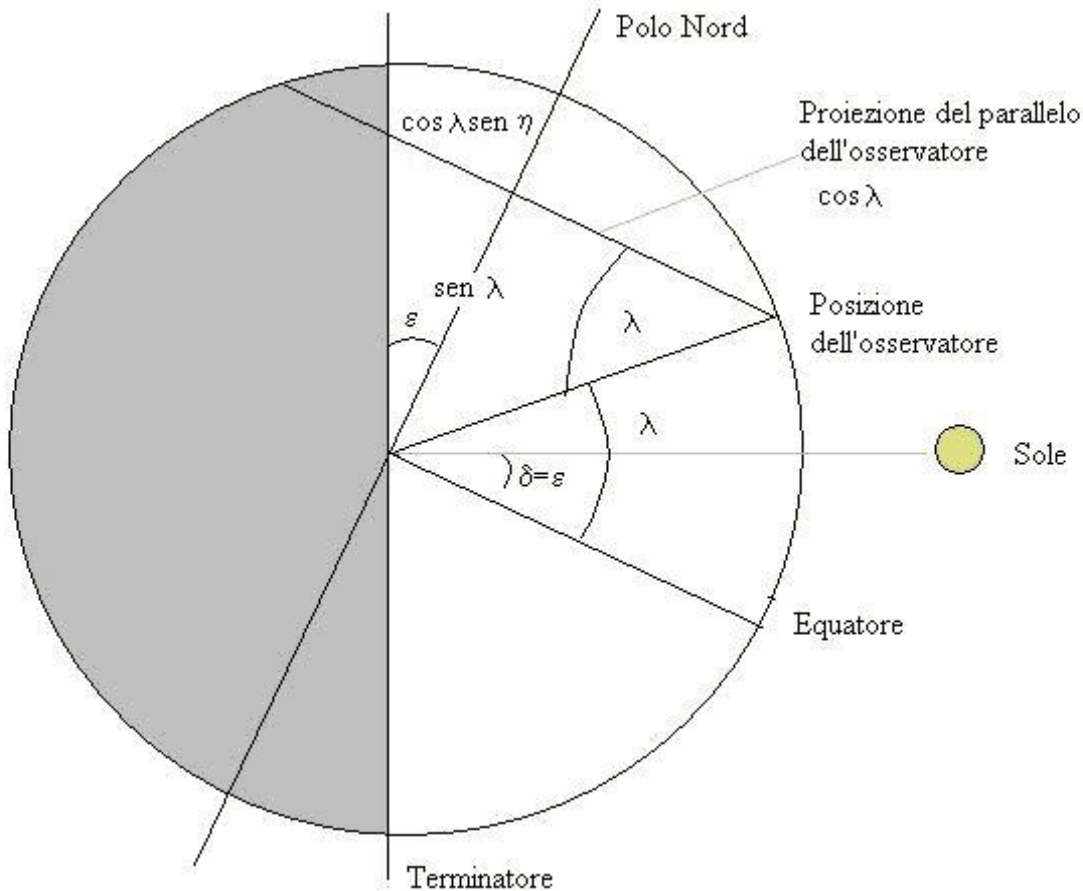

FIGURA 3

**Figura esplicativa per la formula (1).**

Per risalire dalla proiezione del settore circolare del cerchio di latitudine illuminato calcoliamo l'angolo alla circonferenza. Questo angolo η visto dal centro del cerchio di latitudine (in più rispetto all'angolo retto che si ha guardando l'equatore dal centro della Terra e che in figura corrisponde al raggio della sfera in proiezione) è presente anche dalla parte di dietro della figura, e perciò corrisponde a metà del tempo di luce in più delle 12 ore dell'equatore.
L'angolo η è visto dal centro del cerchio di latitudine dell'osservatore; noi ne vediamo la proiezione, cioè il raggio del cerchio di latidine, cos(λ), moltiplicato per il seno di η.

Dalla proiezione ricaviamo anche la relazione cos(λ)·sen(η)/sen(λ)=tan(ε), da cui otteniamo senη=tan(ε)·tan(λ). Infine η è l'arcoseno di questa espressione e lo ritroviamo nella formula (1). Il fattore 2/15 serve a riportare l'angolo in ore equinoziali: ogni ora equinoziale (le nostre ore usuali) corrisponde ad un angolo di 15°. Siccome di frazione illuminata ce n'è anche una parte dietro la figura ecco spiegato il fattore 2.
Le 12 ore additive sono quelle della parte illuminata a destra della proiezione dell'asse Nord-Sud del mondo.
Si è assunto per comodità il raggio della Terra unitario.

L'equazione T(λ) delle ore di luce solare[3] del giorno più lungo in funzione della latitudine è

$$T = 12 + \frac{2}{15} \arcsin(\tan\varepsilon \cdot \tan\lambda) \qquad (1)$$

con gli angoli misurati in gradi. La funzione inversa di tale relazione ci da le latitudini corrispondenti ai giorni di durata 13, 14, 15… ore.

$$\lambda = \arctan\left(\frac{\sin(T_{max} - 12) \cdot 15°/12}{\tan(\varepsilon)}\right) \qquad (2)$$

Tolomeo nella *Geografia*[4] stabilì i paralleli fondamentali in modo che passando dall'uno all'altro la durata massima del giorno variasse di 15 minuti per i primi 14 paralleli e poi spaziati di mezzora e un'ora. Tolomeo, inventore dell'astrolabio, aveva fondato la trigonometria ed aveva pubblicato le tavole, ma nelle biblioteche europee del X secolo Tolomeo non c'era più.
Gerberto si rifà all'*Astronomia* di Marziano Capella, autore latino tardo che aveva semplificato i contenuti dell'Almagesto troppo complicati per una vasta diffusione.
Gerberto ha modo anche di correggere degli errori (di trascrizione) che compaiono nei manoscritti di Marziano sulla durata minima del giorno per l'Ellesponto (durata minima che è sempre complementare a 24 ore della durata massima).
Ai fini di comprendere meglio la Geografia antica vale la pena computare tutti paralleli fondamentali tolemaici dall'equatore al circolo polare, seguendo le istruzioni di Tolomeo ed usando la funzione (2).

---

[3] Si noti che in questa derivazione non si tiene conto dell'elevazione sul livello del mare né della rifrazione atmosferica. Questa fa in modo che il Sole si veda anche quando esso si trova già o ancora al di sotto dell'orizzonte per cui la durata del giorno aumenta di $\Delta t = \frac{480}{\cos(\lambda)}$ secondi. Cfr. Peter DUFFET SMITH, *Astronomia Pratica con l'uso del calcolatore tascabile,* Sansoni, Firenze 1983, p.45.
[4] Claudio TOLOMEO, *Geografia*, Libro I, capitolo XXIII.

**Paralleli fondamentali secondo Tolomeo**

| Durata massima del dì (ore e frazioni decimali) | Latitudine oggi (°) ε=23,43° | Latitudine al tempo di Tolomeo (°) con ε=23,833° |
|---|---|---|
| 12,00 | 0,00 | 0,00 |
| 12,25 | 4,32 | 4,24 |
| 12,50 | 8,58 | 8,42 |
| 12,75 | 12,75 | 12,51 |
| 13,00 | 16,76 | 16,46 |
| 13,25 | 20,60 | 20,24 |
| 13,50 | 24,24 | 23,83 |
| 13,75 | 27,65 | 27,21 |
| 14,00 | 30,85 | 30,37 |
| 14,25 | 33,82 | 33,31 |
| 14,50 | 36,57 | 36,04 |
| 14,75 | 39,11 | 38,57 |
| 15,00 | 41,45 | 40,90 |
| 15,25 | 43,60 | 43,05 |
| 15,50 | 45,59 | 45,04 |
| 16,00 | 49,09 | 48,54 |
| 16,50 | 52,05 | 51,51 |
| 17,00 | 54,56 | 54,04 |
| 17,50 | 56,69 | 56,18 |
| 18,00* | 58,50* | 58,01* |
| 19,00 | 61,36 | 60,89 |
| 20,00 | 63,42 | 62,98 |

Si noti che sulle edizioni antiche della *Geografia*, i valori delle latitudini riportati sono quelli della terza colonna, che io ho calcolato inserendo al posto di ε il valore di 23°50' dei tempi di Tolomeo.
Oggi il valore dell'inclinazione dell'asse terrestre sul piano dell'orbita (dell'eclittica) è ε=23°27'. Questa piccola differenza rende ragione dello scarto con i dati computati da Tolomeo con le sue tavole trigonometriche.
*La riga contrassegnata con l'asterisco non è presente in Tolomeo, ma è la latitudine a cui si riferisce Gerberto nella redazione dell'*Horologium* più settentrionale.
Possibile indizio, questo, che Adamo si trovasse più a Nord di Reims, e che per tutte le posizioni intermedie tra l'Ellesponto (40° 55') e i 58° (dati su ε di Tolomeo) Gerberto sottintendesse che fosse sufficiente fare un'interpolazione lineare.

**4. L'incremento della durata del giorno a partire dal solstizio invernale**

Gerberto, ancora una volta, ha a disposizione il testo di Marziano Capella, di cui riporta lo stralcio nella lettera ad Adamo: "Bisogna sapere che il giorno aumenta dal giorno più corto in modo tale che nel primo mese aumenta di 1/12 di quel tempo che è aggiunto in estate (al giorno più lungo). Nel secondo mese 1/6. Nel terzo ¼, e nel quarto ancora ¼. Nel quinto 1/6 e nel sesto 1/12."
Ecco perché è sufficiente avere la durata massima del giorno per computare un almanacco.
Gerberto si riferisce a questa come ad un'ipotesi di Marziano, che egli segue al posto di altre che affermano che da un mese all'altro la lunghezza del giorno aumenti in modo costante. Gerberto segue questa ipotesi perché evidentemente ha osservato e misurato le diverse lunghezze del giorno nei mesi dell'anno.

5. **Computo analitico degli *Horologia* di Gerberto**

La formula della durata del giorno per una data precisa dell'anno è molto simile alla formula (1). E' infatti sufficiente sostituire ad ε il valore della declinazione δ del Sole nel momento in cui si vuole calcolare la durata del dì.
Dalla trigonometria sferica, assumendo che il Sole si muova sull'eclittica con velocità angolare uniforme (un'approssimazione largamente sufficiente ai nostri scopi) ricaviamo la declinazione, sapendo che essa si annulla in corrispondenza dei nodi dell'eclittica, cioè dove essa interseca l'equatore celeste: agli equinozi.

$$\delta = \arcsin(\sin\varepsilon \cdot \sin(30° \cdot t_{mesi})) \qquad (3)$$

Ponendo la formula (3) nella (1) al posto di ε si ottiene

$$T = 12 + \frac{2}{15}\arcsin(\tan\lambda \cdot \tan(\arcsin(\sin\varepsilon \cdot \sin(30° \cdot t_{mesi})))) \qquad (4)$$

Usando questa formula possiamo ricompilare al computer gli *horologia* gerbertiani:

**Orologio per l'Ellesponto**

| Mesi trascorsi dal solstizio invernale | Durata del giorno (ore) | Dati calcolati secondo Marziano | Dati di Gerberto (dì) | Gerberto (notte) |
|---|---|---|---|---|
| 0 | 9,00 | 9 | 9 | 15 |
| 1 | 9,48 | 9 ½ | 9 | 15 |
| 2 | 10,63 | 10 ½ | 10 ½ | 13 ½ |
| 3 | 12,00 | 12 | 12 | 12 |
| 4 | 13,37 | 13 ½ | 13 ½ | 10 ½ |
| 5 | 14,52 | 14 ½ | 14 ½ | 9 ½ |
| 6 | 15,00 | 15 | 15 | 9 |
| 7 | 14,52 | 14 ½ | 14 ½ | 9 ½ |
| 8 | 13,37 | 13 ½ | 13 ½ | 10 ½ |
| 9 | 12,00 | 12 | 12 | 12 |
| 10 | 10,63 | 10 ½ | 10 ½ | 13 ½ |
| 11 | 9,48 | 9 ½ | 9 | 15 |
| 12 | 9,00 | 9 | 9 | 15 |

**Orologio per coloro il cui giorno più lungo vale 18 ore equinoziali.**

| Mesi trascorsi dal solstizio invernale | Durata del giorno (ore) | Dati calcolati secondo Marziano | Dati di Gerberto (dì) | Gerberto (Notte) |
|---|---|---|---|---|
| 0 | 6,00 | 6 | 6 | 18 |
| 1 | 7,10 | 7 | 6 | 18 |
| 2 | 9,43 | 9 | 9 | 15 |
| 3 | 12,00 | 12 | 12 | 12 |
| 4 | 14,57 | 15 | 15 | 9 |
| 5 | 16,90 | 17 | 17 | 7 |
| 6 | 18,00 | 18 | 18 | 6 |
| 7 | 16,90 | 17 | 17 | 7 |
| 8 | 14,57 | 15 | 15 | 9 |
| 9 | 12,00 | 12 | 12 | 12 |
| 10 | 9,43 | 9 | 9 | 15 |
| 11 | 7,10 | 7 | 6 | 18 |
| 12 | 6,00 | 6 | 6 | 18 |

I dati forniti da Gerberto sono in buon accordo con questi andamenti, e con la precisione di mezz'ora con cui si presentano i dati.

Gerberto fornisce i dati a coppia. Come si vede dalle due tabelle ci sono dei valori che si ripetono, nel computo analitico: 0 e 12 mesi corrispondono alla stessa data, l'istante del solstizio, e 1 ed 11 corrispondono ad un mese dopo (1) e uno prima (11) del solstizio invernale dell'anno successivo. Gerberto include questi 4 dati in quello di Gennaio-Dicembre.

Per Gerberto Giugno e Luglio costituiscono due situazioni simmetriche, così come Maggio e Agosto, Aprile e Settembre, Marzo ed Ottobre, Febbraio e Novembre e Gennaio e Dicembre.

Come dire che la scelta di presentare questi dati a coppia corrisponda a quella di piazzare il Solstizio esattamente in mezzo tra Dicembre e Gennaio, cioè il primo del mese di Gennaio. Mentre ai suoi tempi il solstizio d'inverno, a causa del ritardo accumulato dal calendario giuliano sui fenomeni astronomici, cadeva attorno al 13 Dicembre.

La tabella delle durate della notte sono ottenute da Gerberto come complemento a 24 ore.

Come mai i dati calcolati da Gerberto differiscono da quelli secondo i dettami di Marziano Capella, ed inoltre sono in minore accordo con i dati calcolati analiticamente?

La scelta fatta da Gerberto di considerare i mesi dell'anno a coppia, tra le variabili indipendenti del calcolo, fa si che egli abbia trascurato nel dato dei due mesi Dicembre-Gennaio l'incremento di 1/12 previsto per il primo mese di distanza temporale dal solstizio.

**Bibliografia**